\documentclass[cits]{PoS}

\title{Kinematic studies of the IDV quasar 0917+624}

\ShortTitle{Kinematic studies of 0917+624}

\author{\speaker{Simone Bernhart}\\
        Max-Planck-Institut f\"ur Radioastronomie, Bonn, Germany\\
        E-mail: \email{simone@mpifr-bonn.mpg.de}}

\author{Thomas P. Krichbaum\\
        Max-Planck-Institut f\"ur Radioastronomie, Bonn, Germany\\
        E-mail: \email{tkrichbaum@@mpifr-bonn.mpg.de}}

\author{Lars Fuhrmann\\
        Max-Planck-Institut f\"ur Radioastronomie, Bonn, Germany\\
        E-mail: \email{lfuhrmann@mpifr-bonn.mpg.de}}

\author{Alexander Kraus\\
        Max-Planck-Institut f\"ur Radioastronomie, Bonn, Germany\\
        E-mail: \email{akraus@mpifr-bonn.mpg.de}}

\abstract{For more than a decade from the late 1980s to the late 1990s,
the intraday variable (IDV) quasar 0917+624 was known to be strongly
variable with amplitude variations of 10-20\% within 0.8 to 1.6
days. During that period it showed faster variability in the
polarized flux and the polarization angle than in total
intensity. However, since 2000 the IDV of this source has
almost ceased, attaining a variability index of only less than a few
percent. If interstellar scintillation was responsible for the
IDV in the past, one possible explanation for the sudden change of the
variability mode could be an intrinsic change in the morphology of the
source, such as an increase in the size of the structural components
that are responsible for the scintillation. New jet components
appearing at the jet base could lead to a temporarily larger size of
the scintillating component when compared to the scattering size of
the interstellar medium, causing quenched scintillation. In order to
test this scenario, VLBI observations of 0917+624 from 1999 to 2005
have been analysed. Here, we summarize first results and discuss them
in the framework of the quenched scintillation model.}

\FullConference{The 8th European VLBI Network Symposium\\
                September 26-29, 2006\\
                Toru\'n, Poland}

\begin{document}
\section{Introduction}
\begin{figure}[htbp]
  \centering
\hspace{-1cm}
  \begin{minipage}[c]{6 cm}
    \includegraphics[bb=80 60 570 685,width=0.75\textwidth,angle=-90,clip]{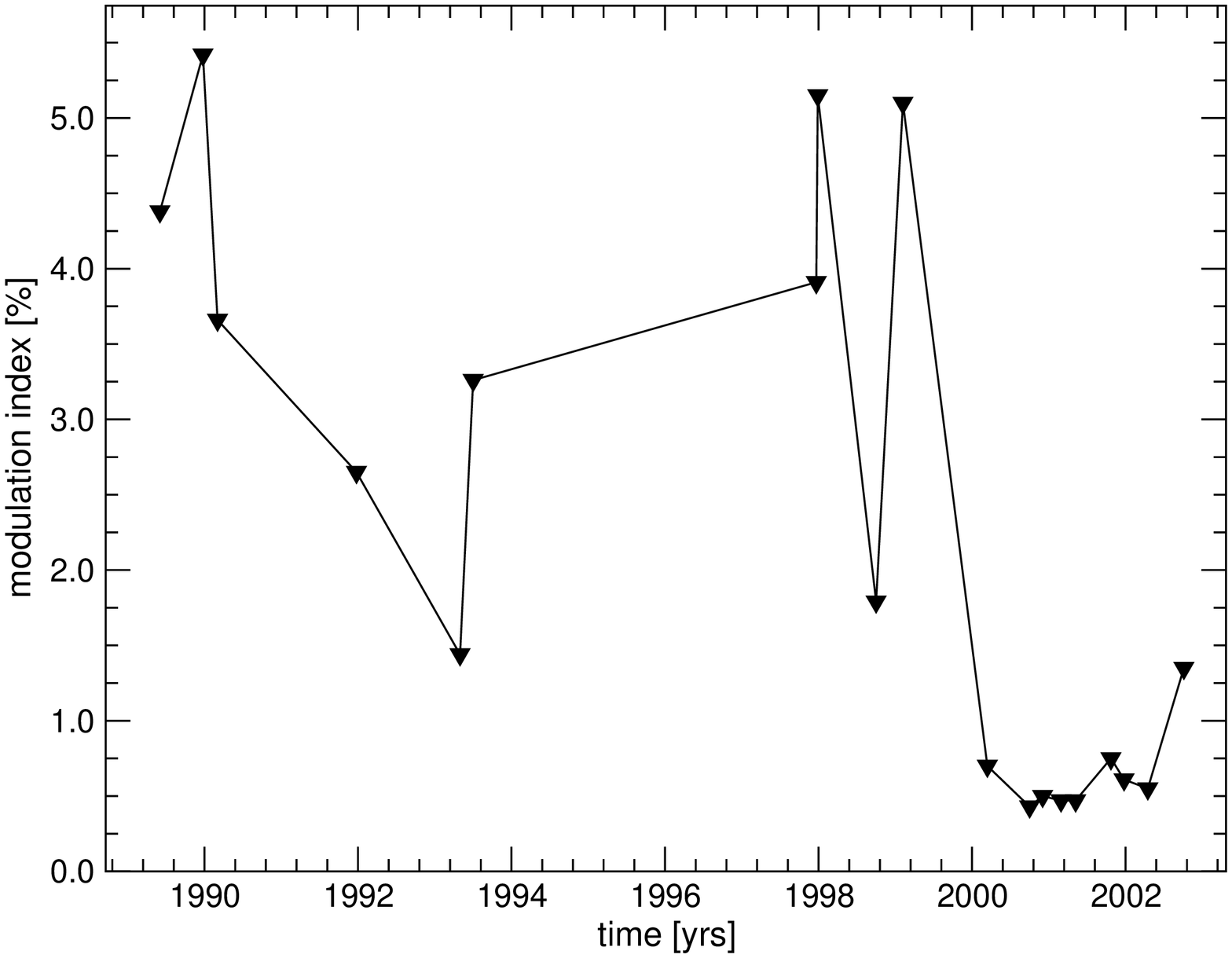}
  \end{minipage}
\hspace{2cm}
  \begin{minipage}[c]{3.5 cm}
  \includegraphics[bb=0 108 419 695,width=0.9\textwidth,clip]{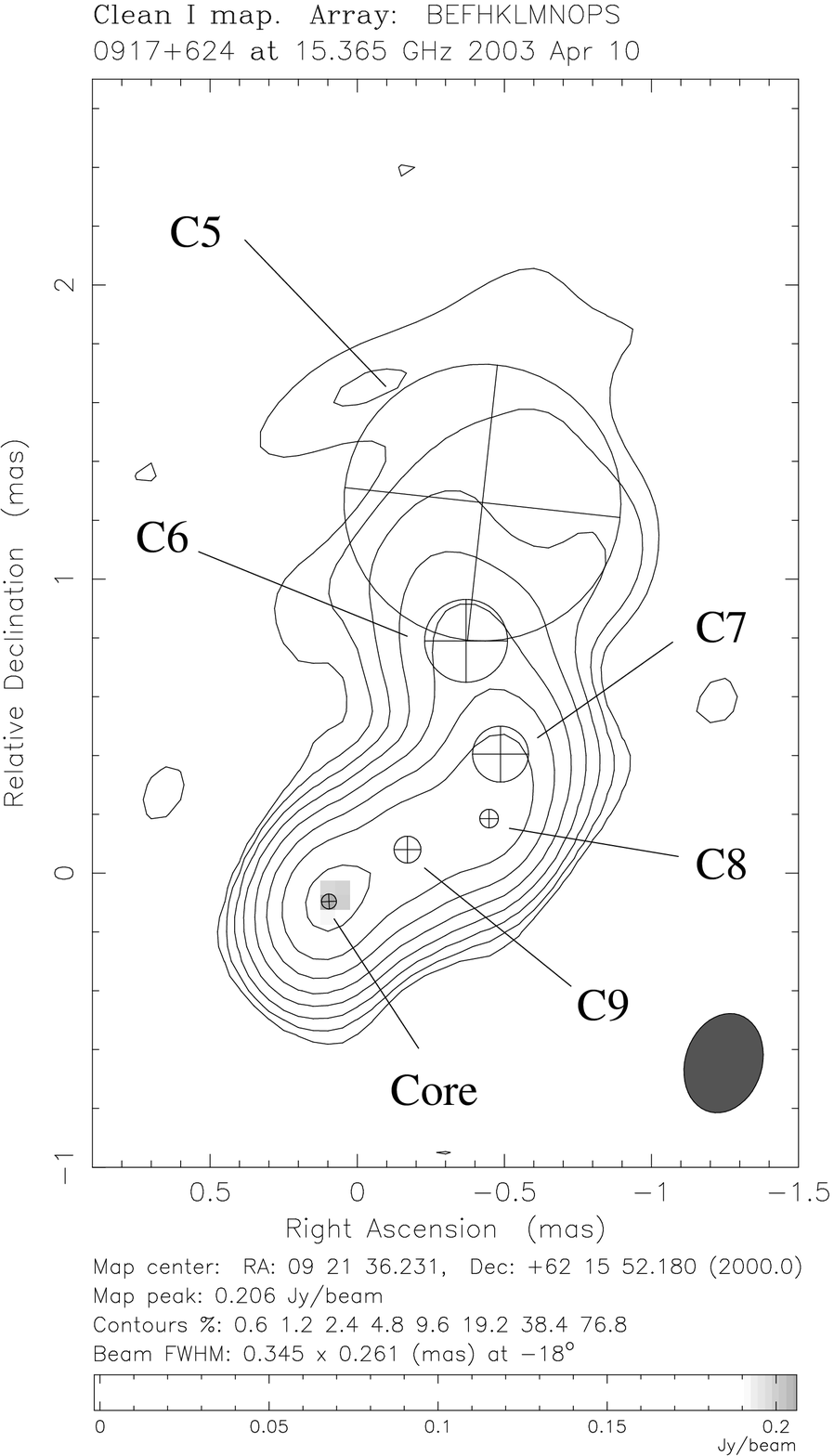}
  \end{minipage}
\caption{{\it Left}: Modulation index of 0917+624 versus time from 1989 to 2002; {\it Right:} VLBI map of 0917+624 at 2\,cm observed in April 2003, beam 0.345$\times$0.261\,mas, beam position angle $-18^{\circ}$; the map is a representative example for the component identification in all epochs. }
 \label{fig1}
\end{figure}

Extragalactic compact flat-spectrum radio sources are known to be
highly variable and about 30\% of them show intra-day variability
(IDV) (e.g., \cite{1},\cite{2},\cite{3}). 0917+624 (z=1.446) is a
type-II IDV quasar. (See \cite{4} for a more detailed description of
IDV). The source used to be strongly variable on timescales of 0.8 to
1.6 days with modulation indices {\it m}\footnote{$m[\%]=\sigma_{S}/<S>$,
where $\sigma_S$ is the rms flux density} of 3 to 5\% at
cm-wavelengths from 1986 to 1998. Moreover, in the polarized flux, as
well as in the polarization angle, even faster variability was
detected (\cite{5}); the polarized flux usually being anti-correlated
with the total flux (\cite{6}). The source structure on
milliarcsecond scales reveals a compact core with a jet pointing in
northerly direction (see right panel of Figure\,\ref{fig1}), whereas
on arcsecond scales the source is mainly pointlike showing a jet-like
feature 4\,mas to the south-west (e.g.,\,\cite{7}).

In September 1998, Kraus et al. (\cite{8}) discovered that the rapid
variability had stopped and instead a slow, monotonic increase of 7\%
in total flux density was observed during the 5-day observing
session. Observations made in February 1999 revealed that 0917+624 was
varying again on a time scale of 1.3 days. However, since September
2000 the variability of the source has ceased again, showing only
moderate modulation indices of the order of $\sim0.5\%$
(see\,\cite{9}, \cite{10}), and to date the source has not restarted
its former activity.

The question whether the origin of IDV is source extrinsic, i.e. due
to scintillation in the interstellar medium (ISM), or intrinsic,
remains unanswered. It is expected, though, that IDV blazar cores
feature micro- or even nanoarcsecond angular sizes and therefore must
scintillate through the ISM. Whether it be extrinsic or intrinsic,
possible explanations for the reduced IDV activity could be either
that the scintillating medium has changed (e.g., strength of
turbulence, distance), or the flux of the scintillating component(s)
has now decreased and become less dominant. Another possibility is the
precession of the footpoint of the jet, which has been proposed in the
cases of 0716+714 \cite{11} and 1803+784 \cite{12}, i.e. the size of
the VLBI core changes with time due to precession.

Furthermore, the earlier slow-down of 0917+624 in September 1998 was
interpreted by Kraus et al. (\cite{8}) as probably due to either a
disappearance of the scintillating compact component or an increase in
its angular size, which could be caused by the ejection of a new jet
component (see also\,\cite{13}). The ejection could temporarily lead
to a core size exceeding the Fresnel scale defined by the scattering
medium. Consequently, only strongly quenched scintillation is
observed. When the component moves further down the jet, it will
separate from the core at some point whereby the size of the
scintillating component decreases again. At that time, the variability
is expected to reappear. Given that the variability ceased again in
2000, even more than one component should have been emitted since. In
order to test this scenario, new VLBI observations of 0917+624 were
carried out at 6\,cm, 2\,cm and 1.3\,cm at 4 epochs, covering the
period between December 2001 and April 2003. Here, we present the
results of the analysis of our 2-cm data complemented with additional
2-cm data, including 3\,VLBI epochs observed between November 1999 and
November 2000 (\cite{9}, \cite{13}) and 6\,epochs from the
MOJAVE\footnote{Monitoring of Jets in Active galactic nuclei with VLBA
Experiments} (\cite{14}) and 2cm Survey (\cite{15}) programs observed
between June 2002 and August 2006. A more detailed description of the
observations and the data reduction will be given in Fuhrmann et
al. in prep.

\section{Results and discussion}
\subsection{Kinematic analysis}
\begin{figure}[h]
  \centering
  \begin{minipage}{10 cm}
  \includegraphics[bb=0 0 720 475,width=0.85\textwidth,clip]{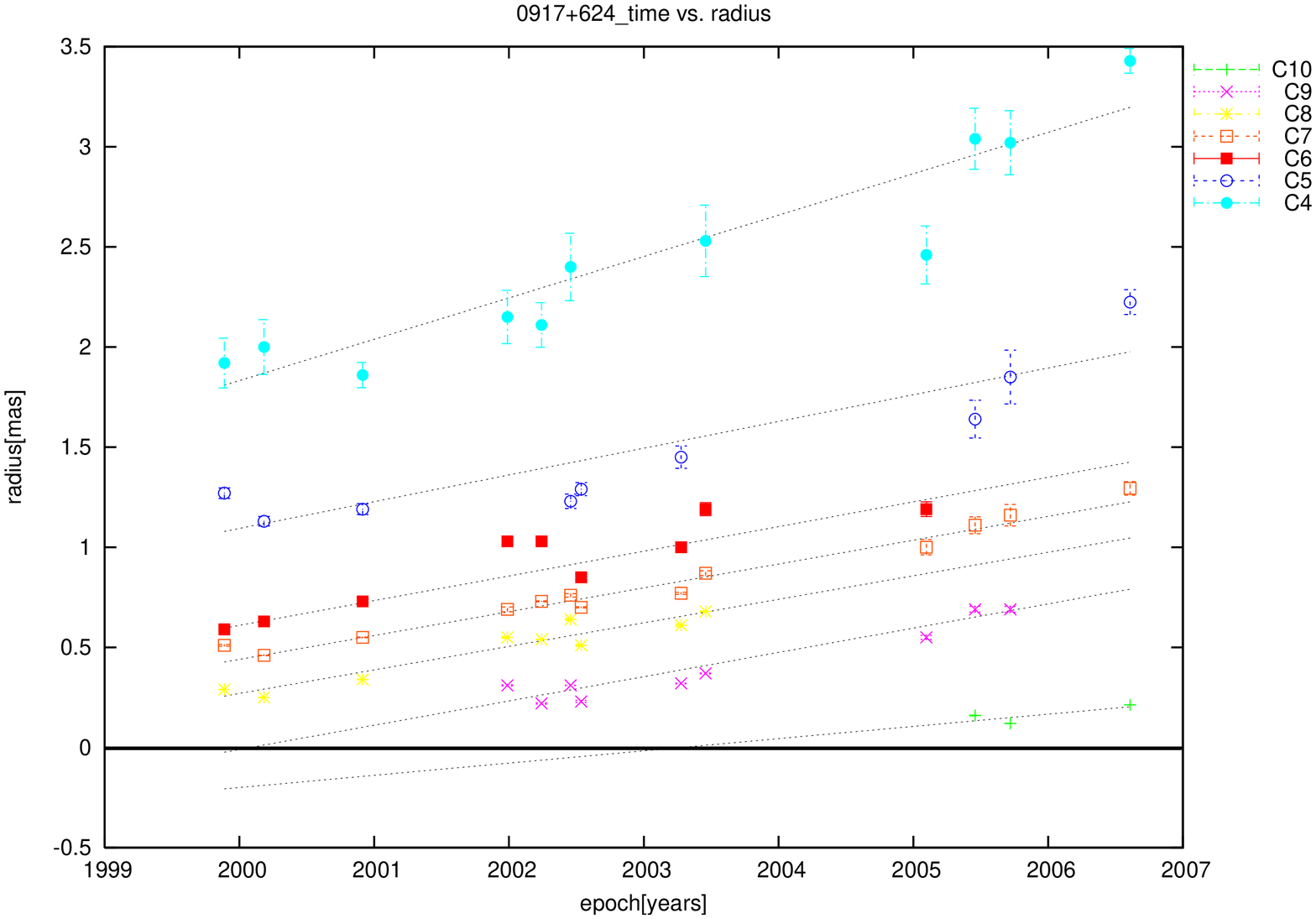}
  \end{minipage}
\hspace{0.5cm}
  \begin{minipage}{4 cm}
\includegraphics[bb=115 380 290 510,width=\textwidth,clip]{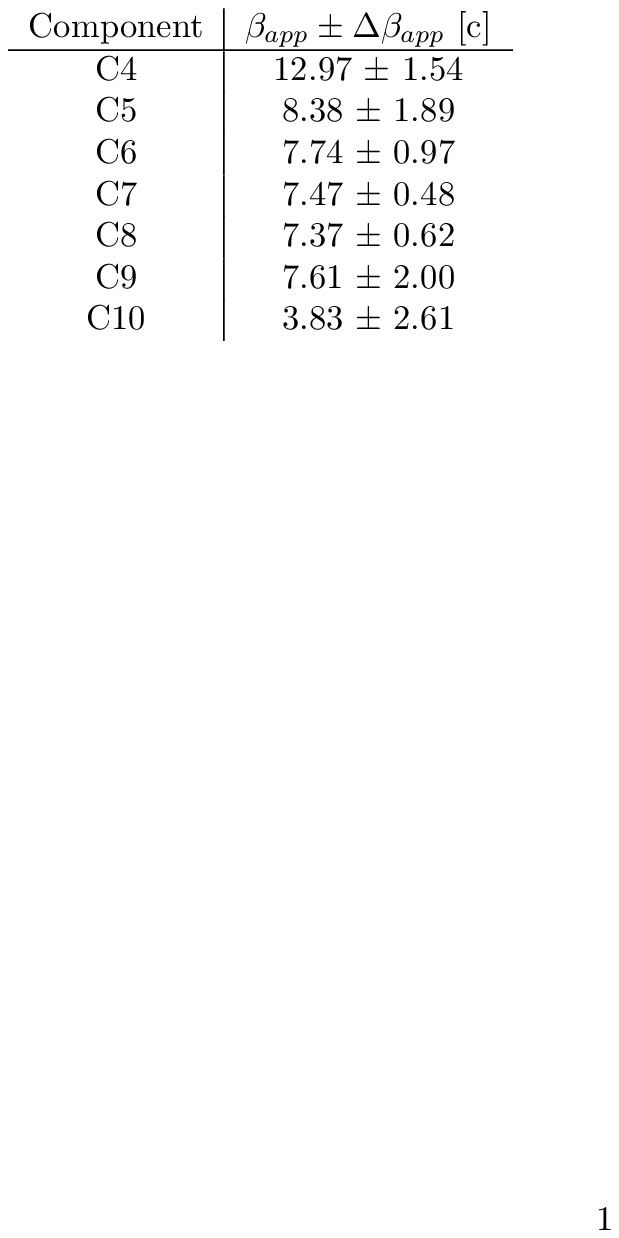}
  \end{minipage}
  \caption{Core separation of the individual modelfit components plotted versus time; the solid lines are linear fits to the path of each component; from the slope the apparent velocities are derived which are listed in the Table.}
 \label{fig2}
\end{figure}

To investigate the jet kinematics of 0917+624 we fitted a number of
circular Gaussian components to the calibrated visibilities in order
to describe the source with as few parameters as possible. For each
epoch we cross-identified individual model components along the jet
using their distance from the VLBI core, which is assumed to be
stationary, the flux density and size. The right panel of
Figure\,\ref{fig1} shows a representative map of the component
identification. The higher numbered components are the ones that were
expelled last. By means of a graphical analysis, which is presented in
the left panel of Figure\,\ref{fig2}, we obtained an adequate
identification scheme for the kinematics in the jet of 0917+624. This
scenario consists of 10 superluminal components in total, moving
linearly away from the core, the innermost 7 of which are displayed
here.

\begin{figure}
\hspace{1.5cm}
   \includegraphics[bb=550 50 80 800,width=0.4\textwidth,angle=90,clip]{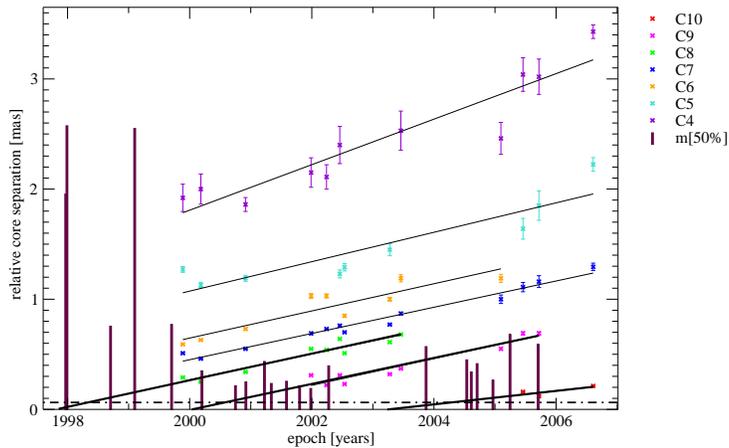} 
\caption{Core separation plotted with modulation indices (scaled down by 50\%) displayed as bars; the dashed-dotted line indicates the typical scattering size ($\sim$0.07\,mas) for this source as defined by the ISM (see\,\cite{16})}
 \label{fig3}
\end{figure}

From the slope of the linear fit to the path of the components, the
individual apparent velocities can be calculated. The results are
listed in the Table of Figure\,\ref{fig2}. The inner components C9,
C8, C7, C6, C5 move with an apparent speed of $\beta_{app}$=7.4 to 8.4
(adopting $H_0$ = 73\,kms$^{-1}$\,Mpc$^{-1}$), which is in good
agreement with previous findings (\cite{13}). Moreover, a
back-extrapolation of the fitted line to the time-axis delivers an
approximate ejection date for the respective component (see
Figure\,\ref{fig3}). For component C8, the ejection date lies around
1997.8, which is in the range of earlier results found by Krichbaum et
al.\,\cite{13} and Fuhrmann\,\cite{9}. That event corresponds to a
high level of variability. The modulation index lies at $\sim$5\% and
decreases in the aftermath to $\sim$1.5\%. That was about the time
when the size of the core and the emitted component exceed the typical
scattering size of the medium (indicated by the dashed-dotted line in
Figure\,\ref{fig3}), presumably resulting in strongly quenched
scintillation. Considering the succeeding increase in the modulation
index in early 1999, the behaviour of C8 could in principle confirm
our working hypothesis.

Component C9 was approximately ejected in the beginning of 2000, which
overlaps with a period of very low variability according to the
modulation indices around that time. The data for the last but two
observing epoch reveal that another component (C10), which was not
visible in the earlier epochs, has been emitted moving with an
apparent speed of $\beta_{app}$=3.83$\pm$2.61. The back-extrapolation
yields an approximate ejection date of 2002.3. However, ejection date
and apparent speed are affected by a high uncertainty due to a large
(>50\%) error in the slope of the component path, since only
three-epoch measurements are available for C10 so far.

\subsection{Flux density measurements}

\begin{figure}
  \centering
   \includegraphics[bb=30 40 790 470,width=0.75\textwidth,clip]{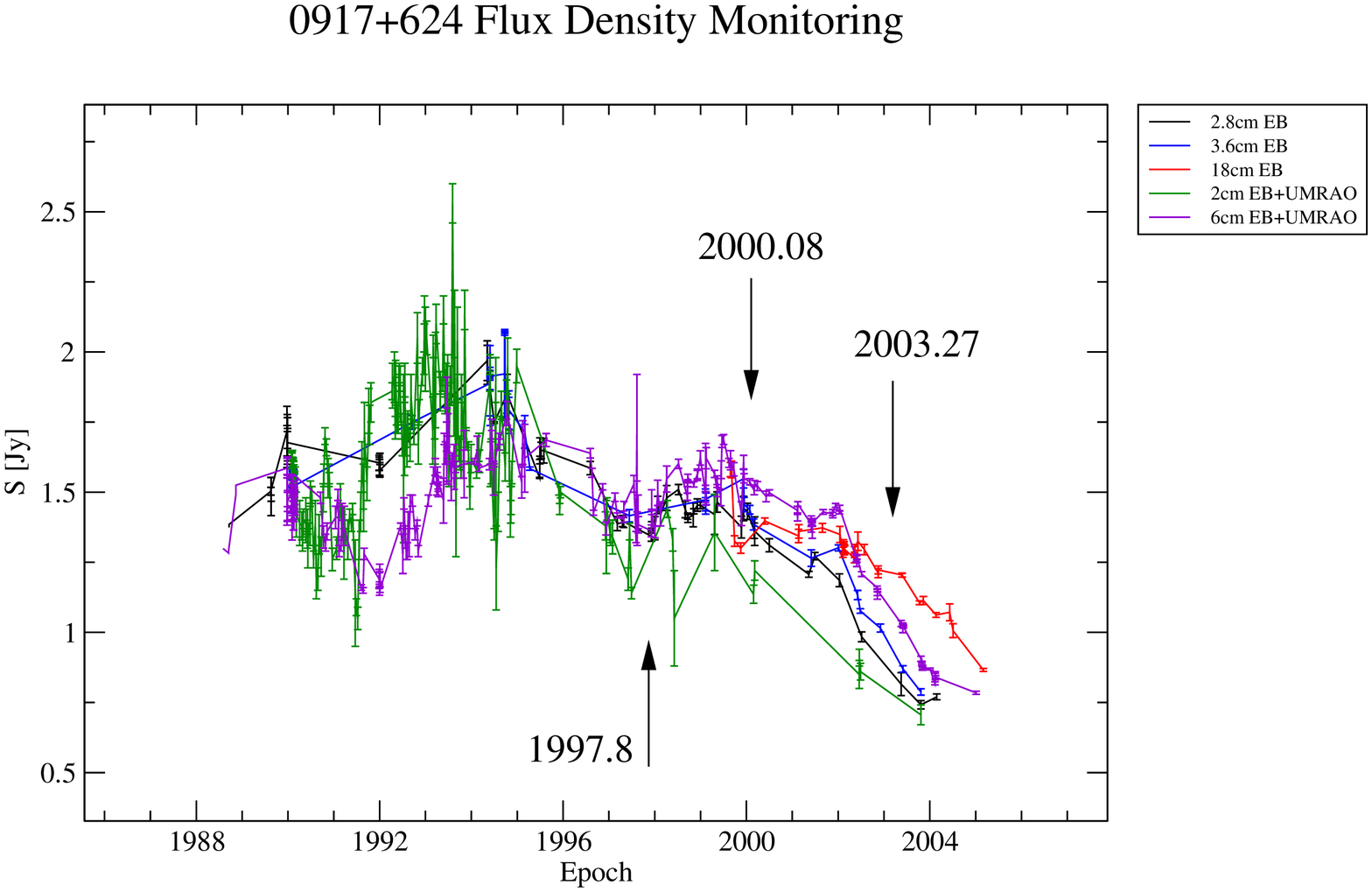}  
\caption{Multi-frequency plot of flux density observations carried out with the Effelsberg radio telescope and extracted from the UMRAO data base; the arrows indicate the ejection dates of new jet components.}
 \label{fig4}
\end{figure}

Theoretical simulations have shown that the ejection of a new VLBI
component can be accompanied by a flux outburst at all
frequencies\,\cite{17}. The flux reaches maximum when the component
becomes optically thin, which depends on the opacity at each
frequency. The flux peak is higher and emerges earlier at higher
frequencies. This general behaviour in flux outburst at various
frequencies is shown in Figure\,\ref{fig4} where long-term lightcurves
of 0917+624 at 7 different frequencies (from 1.3 to 18\,cm) are
displayed. The data were obtained from observations with the
Effelsberg 100-m radio telescope and also complemented by
UMRAO\footnote{University of Michigan Radio Astronomy Observatory}
data. From the slope after a major flare, the time delay between the
highest and lowest frequency can be derived. This gives an
approximation of the period that a newly ejected component needs to
separate from the VLBI core. After this time, the IDV activity is
expected to resume. In the case of 0917+624, the delay is
approximately 0.8\,years (N. Kudryavtseva, priv. comm.) for the large
flare around 2000, which coincides with the ejection date we found for
component C9. This implies that at least one year after this flare, the
source should exhibit a higher level of IDV activity again, which is,
however, not the case according to the modulation indices around 2001
(see\,Figure\,\ref{fig4}). Furthermore, it seems that a burst in flux
that is related to a new component ejection, is not directly related to
the IDV activity of a source, since the modulation index was
comparatively low at the time of the flare.

\section{Conclusion}
We found two new VLBI components being ejected around the beginning of
2000 and possibly in early 2003. The gap between the two ejection
dates is too long compared to the time obviously needed for a
component to travel down the jet and separate from the
core. Therefore, the IDV activity should meanwhile have resumed. Since
this is not the case, the hypothesis of a quenched scintillation
scenario is not conclusive. Taking the existing data, we will now
focus on the alternative scenarios, e.g., the analysis of
changes in the flux density of the scintillating component. Since the
VLBI observations were carried out in dual-polarization mode, we will
also have a closer look to the polarization analysis. In addition, we
will work out the spectral properties of the individual jet components
using the three available frequencies. We further plan to
continue the single dish monitoring in order to find out if and when
0917+624 is going to return to its former prominent IDV activity.

\acknowledgments 
We thank the MOJAVE and 2cm Survey programs for providing their
data. We appreciate the use of the flux density monitoring data from
the UMRAO data base. This work has made use of the VLBA, which is an
instrument of the National Radio Astronomy Observatory, a facility of
the National Science Foundation, operated under cooperative agreement
by Associated Universities, Inc. This work is also based on
observations with the 100-m radio telescope of the MPIfR
(Max-Planck-Institut f\"ur Radioastronomie) at Effelsberg.

\end{document}